\documentclass[conference]{IEEEtran}
\IEEEoverridecommandlockouts
\newcommand{\conftitle}{17th International Workshop on Science Gateways (IWSG2025), 17-19 June 2025}
\usepackage{fancyhdr}
\usepackage{cite}
\usepackage{amsmath,amssymb,amsfonts}
\usepackage{graphicx}
\usepackage{tikz}
\usepackage{listings}
\usepackage{algorithm,algpseudocode}
\usepackage{amssymb}
\usepackage{amsmath}
\usepackage{pgfplots}
\usepackage{xcolor}
\usepackage{booktabs}
\usetikzlibrary{arrows,shapes,fit,positioning}
\usetikzlibrary{decorations.pathmorphing}
\usetikzlibrary{decorations.markings}
\def\BibTeX{{\rm B\kern-.05em{\sc i\kern-.025em b}\kern-.08em
    T\kern-.1667em\lower.7ex\hbox{E}\kern-.125emX}}

\pgfplotsset{compat=1.7}
\fancypagestyle{pageStyle}{%
    \fancyhf{}
    \fancyhead[C]{\conftitle}

}
\definecolor{delim}{RGB}{20,105,176}
\definecolor{numb}{RGB}{106, 109, 32}
\definecolor{string}{rgb}{0.64,0.08,0.08}
\lstdefinelanguage{json}{
    showspaces=false,
    showtabs=false,
    breaklines=true,
    postbreak=\raisebox{0ex}[0ex][0ex]{\ensuremath{\color{gray}\hookrightarrow\space}},
    breakatwhitespace=true,
    basicstyle=\fontsize{6.8}{6.8}\ttfamily,
    upquote=true,
    morestring=[b]",
    stringstyle=\color{string},
    literate=
     *{0}{{{\color{numb}0}}}{1}
      {1}{{{\color{numb}1}}}{1}
      {2}{{{\color{numb}2}}}{1}
      {3}{{{\color{numb}3}}}{1}
      {4}{{{\color{numb}4}}}{1}
      {5}{{{\color{numb}5}}}{1}
      {6}{{{\color{numb}6}}}{1}
      {7}{{{\color{numb}7}}}{1}
      {8}{{{\color{numb}8}}}{1}
      {9}{{{\color{numb}9}}}{1}
      {\{}{{{\color{delim}{\{}}}}{1}
      {\}}{{{\color{delim}{\}}}}}{1}
      {[}{{{\color{delim}{[}}}}{1}
      {]}{{{\color{delim}{]}}}}{1},
}
\begin{document}
\title{ArcBERT: An LLM-based Search Engine for Exploring Integrated Multi-Omics Metadata}
\renewcommand\IEEEkeywordsname{Keywords}

\author{\IEEEauthorblockN{Gajendra Doniparthi\IEEEauthorrefmark{1},
Shashank Balu Pandhare\IEEEauthorrefmark{2},
Stefan De{\ss}loch\IEEEauthorrefmark{1} and
Timo M{\"u}hlhaus\IEEEauthorrefmark{3}}
\IEEEauthorblockA{\IEEEauthorrefmark{1}Heterogeneous Information Systems,}
\IEEEauthorblockA{\IEEEauthorrefmark{2}Center for Commercial Vehicle Technology,}
\IEEEauthorblockA{\IEEEauthorrefmark{3}Computational Systems Biology,}
\IEEEauthorblockA{RPTU Kaiserslautern-Landau, Gottlieb-Daimler-Straße 47, 67663 Kaiserslautern, Germany}
\IEEEauthorblockA{gajendra.doniparthi@cs.rptu.de, pandhare@rptu.de, stefan.dessloch@cs.rptu.de, timo.muehlhaus@rptu.de}}
\maketitle
\thispagestyle{pageStyle}
\pagestyle{fancy}
\renewcommand{\headrulewidth}{0pt} 

\begin{abstract}
Traditional search applications within Research Data Management (RDM) ecosystems are crucial in helping users discover and explore the structured metadata from the research datasets. Typically, text search engines require users to submit keyword-based queries rather than using natural language. However, using Large Language Models (LLMs) trained on domain-specific content for specialized natural language processing (NLP) tasks is becoming increasingly common. We present ArcBERT, an LLM-based system designed for integrated metadata exploration. ArcBERT understands natural language queries and relies on semantic matching, unlike traditional search applications. Notably, ArcBERT also understands the structure and hierarchies within the metadata, enabling it to handle diverse user querying patterns effectively.    
\end{abstract}

\IEEEpeerreviewmaketitle
\section{Introduction}
It is widely accepted in the bio-science research community that an effective Research Data Management (RDM) platform based on FAIRness (i.e., data that is findable, accessible, interoperable, and reusable) is essential to improve the reproducibility, visibility, and accessibility of their heterogeneous research data~\cite{RDMinPlantBiology, PlantScienceDataIntegration, HAUG201758, DataManagementStrategies}. The search \& exploration applications play a pivotal role in fully integrating the standardized experimental metadata within and across various RDM eco-systems. They are expected to provide a near real-time consolidated view of the metadata while hiding the heterogeneity of the data model and data exchange methodologies among the participating groups. Exploratory query processing techniques enable scientists to interactively explore the metadata and associated datasets, identify patterns, and generate data-driven insights. Popular open-source endpoint repositories, such as MetaboLights~\cite{MetaboLightsRepo, MetaboLights} and PRIDE~\cite{PridePaper}, enable users to conduct direct free-text metadata searches. They also offer search facets, allowing users to filter results by specific studies or organisms.


As large language models (LLMs) gain popularity in bioscience research, they represent a promising alternative to traditional text search engines for exploring standardized research data~\cite{LLMsInPlantBiology}. Modern language models comprehend search keywords and the semantics behind user queries, allowing for more accurate results. LLMs can potentially transform how we analyze integrated experimental metadata, providing a more powerful and insightful tool than conventional text search engines.

A comprehensive list of large language models (LLMs) trained on extensive biological corpora~\cite{ScientificLargeLanguageModels} exists for the bioscience domain. Domain-specific LLMs such as BioBERT~\cite{BioBERT} and SciBERT~\cite{SciBERT} have demonstrated effectiveness in tasks like question answering and text mining. These models are developed by initially pre-training a BERT~\cite{BERTPaper} model on general domain corpora (for instance, Wikipedia) and further training it on specific domain corpora. BioBERT is fine-tuned with biomedical texts from sources like PubMed and PMC, whereas SciBERT is fine-tuned with scientific literature from Semantic Scholar. Fine-tuning the models on particular biological natural language processing (NLP) tasks leads to a significantly improved understanding of biological terminology.

There are three key aspects to consider when using LLMs to develop a system to explore multi-omics metadata in plant sciences. First, leveraging the RDM solution and its features is essential to building a comprehensive corpus of integrated omics metadata. Second, the model needs to be fine-tuned; this involves using a pre-trained model based on plant science literature and adapting it with the omics metadata corpus to enhance its ability to understand user queries in natural language. Lastly, since the metadata specifications follow a hierarchical structure, the model must grasp these layers to match user queries meaningfully. This paper presents the following contributions.
\begin{itemize}
    \item We revisit our RDM approach (\emph{PLANTdataHUB}) and the existing metadata search application, \emph{ARC Metadata Registry}, describing how our RDM solution structures and integrates the multi-omics metadata.   
    \item We introduce \emph{ArcBERT}, a Sentence-BERT~\cite{SentenceBERT}-based language model for integrated metadata exploration. The model is pre-trained using plant science literature from PubMed~\cite{PubMed} and fine-tuned with integrated omics metadata.
    \item We propose using an indexing mechanism to index the vector embeddings from ArcBERT and leverage the FAISS library for fast nearest neighbor searches.   
    \item We evaluate the effectiveness of using ArcBERT for metadata exploration compared to traditional text search engines.
\end{itemize}
The following section briefly describes our PLANTdataHUB ecosystem and its core components. Sec.~\ref{sec:integration} outlines how the ARC Metadata Registry application connects with multiple dataHUBs and integrates the omics metadata. In Sec.~\ref{sec:model}, we introduce ArcBERT, its pre-training and fine-tuning with the omics metadata from the registry application, and indexing the ARC document and structure embeddings. Lastly, in the experiments section (Sec.~\ref{sec:experiments}), we evaluate ArcBERT against Elasticsearch in finding how effective ArcBERT is for metadata exploration.
\section{PLANTdataHUB}
PLANTdataHUB is a comprehensive FAIR solution designed for managing research data in plant science~\cite{PLANTdataHUB}. In this RDM solution, the units of research output are referred to as \emph{Annotated Research Contexts}, in short, ARCs, which are FAIR digital objects. ARCs are self-contained, interoperable, and reproducible with unique digital object identifier (DOI). The PLANTdataHUB solution leverages a distributed version control system (GitLab) to store and version individual ARCs. It provides flexibility in maintaining each iteration of a laboratory experiment's design-test-repeat cycle, thus avoiding the need for a new customized versioning tool.  
\begin{figure}
    \centering
    \includegraphics[width=\columnwidth]{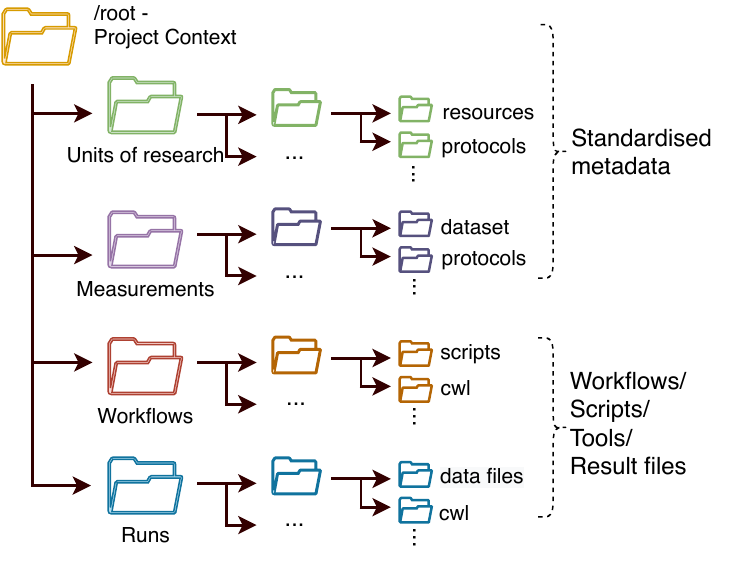}
    \caption{ARC folder specification packaging standard format metadata with workflows, scripts for computational pipelines, and result files/artifacts from workflow executions.}
    \label{fig:arc}
\end{figure}
\subsection{Annotated Research Context}
The core idea behind Annotated Research Contexts (ARC) is to follow a fixed folder and file layout to organize the experimental metadata, measurement data, workflows, software, external files, etc., in a generic and interoperable format to ensure the reproducibility of each experiment within a research project~\cite{ARC}. As shown in Figure.~\ref{fig:arc}, the format of ARC is standardized and flexible to adapt and accommodate simple to complex project scenarios across any omics studies. The ARC specification is designed to be flexible and can accommodate the experimental metadata to be represented in any known metadata standard~\cite{dataplant_community_2023_8302662}.

The initial ARC specification for plant research data, in particular, allows for well-known standards such as ISA. The abstract model for the ISA standard is implemented in hierarchical ISA-JSON file format~\cite{sansone_ISAJson}. The ARC specification also uses a common workflow language for the workflows and scripts to build the computation pipelines. The artifacts from workflows and computational analysis go into their respective run folders. The ARC can also be compiled to other packing methods, such as RO-Crate~\cite{PackagingRO-Crate}.
\subsection{DataHUBs}
In the PLANTdataHUB model, centralized repositories are known as \emph{DataHUBs}. DataHUBs can be deployed on-premise or remotely, depending on the research groups' data sharing and security policies. They offer versioning of ARCs and make it easy to search and discover through standardized API access. Additionally, PLANTdataHUB provides various supporting tools to enable easy curation, collaboration, and maintenance of ARCs on-premise and in remote dataHUBs.
\section{Integrating Metadata} \label{sec:integration}
\emph{ARC Metadata Registry} is a cloud-native application for integrated search and analysis of experimental metadata within the plant science community. The registry application within the PLANTdataHUB solution is designed to integrate multiple DataHUBs at once and provide a consolidated real-time view of the data from the top layers of the ARCs from across DataHUBs. Most importantly, users can search and explore data from different ARCs and across different DataHUBs simultaneously, thus providing a platform for integrated metadata exploration. ARC Metadata Registry exposes a range of secure APIs to access the standardized metadata, i.e., the existing version allows for ISA \& RO-Crate metadata (that uses Schema.org annotations) serialized in JSON \& JSON-LD, respectively.
\begin{figure}[htp]
    \centering
    \includegraphics[width=\columnwidth]{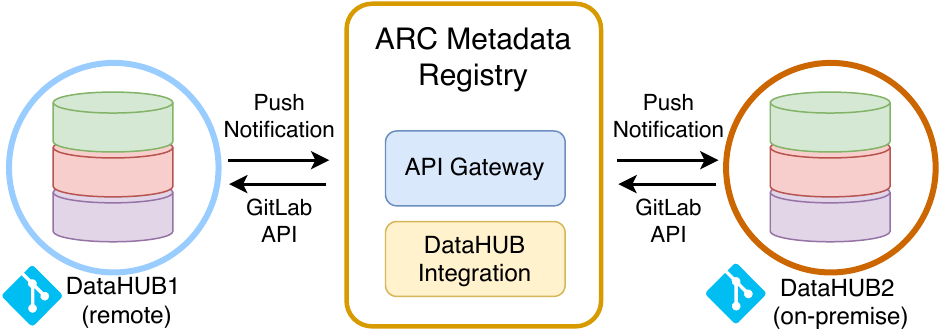}
    \caption{ARC Metadata Registry integrated with multiple on-premise and remote DataHUBs hosting ARCs.}
    \label{fig:hubintg}
\end{figure}
\begin{figure*}[htp]
    \centering
    \includegraphics[width=\textwidth]{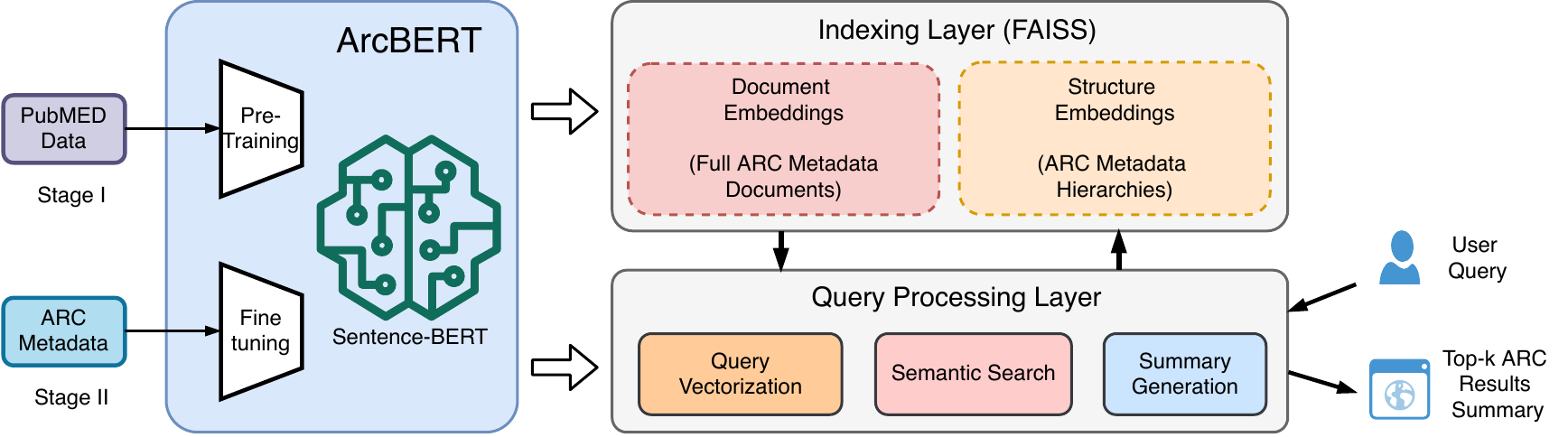}
    \caption{ArcBERT architecture showcasing the Sentence-BERT model, the indexing layer and the query processing layer.}
    \label{fig:pipeline}
\end{figure*}

HUB Integration is a crucial registry application service that integrates multiple DataHUBs and instantly downloads experimental metadata from ARCs. Each participating DataHUB requires a one-time setup to establish a two-way data exchange with the ARC Metadata Registry as shown in Fig.~\ref{fig:hubintg}. Once set up, the ARC Metadata Registry receives real-time push notifications from DataHUBs upon every ARC repository update from the GitLab, and the HUB Integration service downloads and processes the latest changes. There are two essential aspects to integrating DataHUBs hosting ARCs. The first aspect is generating real-time notifications whenever a user updates an ARC. For instance, changes to an existing workflow in an ARC, changes to the administrative metadata, etc. The second aspect is accessing ARCs from the DataHUBs using APIs while abstracting the mode of communication.   

The Web UI service maintains web pages for users to search and explore the experimental metadata.
\section{Model Approach}\label{sec:model}
Large Language Models (LLMs) are built on Transformer architectures and typically contain hundreds of millions to billions of trainable parameters. They are trained on extensive textual corpora, which enables them to excel in understanding natural language. These models can also be fine-tuned for specific domains and disciplines. For instance, Bidirectional Encoder Representations from Transformers (BERT)~\cite{BERTPaper} models are commonly used for tasks such as classification, named entity recognition (NER), and summarization. In contrast, Generative Pretrained Transformers (GPT)~\cite{DBLP:conf/nips/BrownMRSKDNSSAA20} models are primarily utilized for text generation and translation.

A crucial step in adapting any sentence-level semantic model involves transforming text input into dense vector representations that capture semantic meaning. Sentence-BERT~\cite{SentenceBERT}, for example, converts each sentence into a fixed-length vector, allowing for semantic similarity measurement using cosine distance. Sentences that express similar ideas are mapped to points that are close together in this vector space, even if the surface wording differs. When fine-tuned with integrated multi-omics metadata, the model can understand user queries in natural language and provide semantically similar query results, unlike traditional text search engines that rely on keyword matching. 
\subsection{ArcBERT}
We present ArcBERT, a domain-specific semantic search system based on the Sentence-BERT architecture as shown in Figure.~\ref{fig:pipeline}. This system is designed for Annotated Research Contexts (ARCs) that contain highly structured metadata from various omics studies. To efficiently retrieve relevant experimental metadata based on semantic similarity, we create a dense semantic vector index using FAISS, which enables rapid retrieval of the top-k most semantically similar ARC metadata for any given input query~\cite{SimilarityDetectorUsingFAISS}. The model and indexing system are designed to maintain semantic similarity and preserve the hierarchical relationships within the metadata, ensuring meaningful query matching across the different metadata layers.
\subsubsection{Pre-training}
We developed a domain-specific training corpus using PubMed, a widely utilized public digital archive, to train our model with a meaningful representation of the language found in plant science literature. We selected English articles published between January 2000 and December 2023 that contained the keywords “plant science” or “plant biology” in either the title or abstract. The extracted dataset was then cleaned and organized into a list of JSON objects, with each object representing one article. Before training the Sentence-BERT model, we concatenated relevant fields from each article—such as the title, abstract, methodology, and results—into a single string.
\subsubsection{Fine-tuning}
We refine the model using layers from the metadata, field-specific annotations, and nested relationships that accurately represent multi-omics data in practice. The ARC specification provides detailed information about the broader project context, key contributors, study designs, assay types, and workflows. By fine-tuning the model with multi-omics metadata structured according to the ARC specification, we can adapt its embeddings to reflect the underlying semantics and context. We leverage the snapshot of integrated multi-omics data directly from the ARC Metadata Registry application. The dataset is cleaned and flattened into plain text as shown in Listing~\ref{lst:arcdata}, combining the elements from its title, description, studies, assays, contributors, and any available ontological references into the field "search\_text". 
\subsubsection{Document Embedding}
After fine-tuning the ArcBERT model, every ARC document in the dataset is flattened into a respective "search\_text" field, which is then transformed into a fixed-length embedding vector. However, performing similarity comparisons between a query vector and the model's embeddings becomes computationally expensive as the number of embeddings increases, rendering brute-force methods inefficient. We utilize FAISS, a library designed for fast nearest-neighbor searches over dense vectors, to speed up the query processing. FAISS offers various indexing structures; for this purpose, we use the $IndexFlatIP$ type, which calculates inner products between vectors. Since the embeddings are normalized to unit length during preprocessing and fine-tuning, these inner products correspond to cosine similarity scores—a widely used metric for semantic comparison at the sentence level. When a user query is submitted, it is processed through the ArcBERT model to generate an embedding in the same vector space. This embedding is then compared against the FAISS indices to identify the most relevant matches based on cosine similarity.
\begin{lstlisting}[language=json, caption=A snippet of the training dataset for fine-tuning the model with the project information flattened into the string "search\_text", label=lst:arcdata]{}

    "title": "Systems-wide investigation of responses to moderate and acute high temperatures in the green alga Chlamydomonas reinhardtii.",
    "description": "Algae cultures were grown mixotrophically (TAP). After 24h of 35/40 degree celsius, the cells were shifted back to room temperature for 48h. 'omics samples were taken.",
    "people": ["Ningning Zhang (Donald Danforth Plant Science Center)"],
    "publications": "Systems-wide analysis revealed shared and unique responses to moderate and acute high temperatures in the green alga Chlamydomonas reinhardtii.",
    "studies": ["HeatstressExperiment"],
    "assays": ["Proteomics", "Transcriptomics", "Metabolomics", "Growth"],
    
    "search_text": "Systems-wide investigation of responses to moderate and acute high temperatures in the green alga Chlamydomonas reinhardtii. Algae cultures were grown mixotrophically (TAP). After 24h of 35/40 degree celsius, the cells were shifted back to room temperature for 48h. 'omics samples were taken. Ningning Zhang (Donald Danforth Plant Science Center) Systems-wide analysis revealed shared and unique responses to moderate and acute high temperatures in the green alga Chlamydomonas reinhardtii. HeatstressExperiment Proteomics Transcriptomics Metabolomics Growth"
\end{lstlisting} 
\subsubsection{Structure Embedding}
The flattened string for each ARC document provides a comprehensive summary. However, relying on it as a single block during retrieval restricts the system’s ability to highlight specific metadata elements corresponding to a user's query. In scientific data analysis, users often focus on particular layers of metadata, such as a particular study or assay name, data stewards or institutions, and experimental parameter values pertinent to a specific buffer condition.

To enhance search capabilities, the ARC metadata is broken down into smaller, more semantically meaningful segments called "chunks." Each chunk corresponds to a distinct metadata category—such as studies, assays, parameters, individuals, or publications—and is processed independently. This approach allows our system to match entire documents and pinpoint the exact section within the ARC hierarchy that aligns with the user’s query. For example, the natural language query \emph{drought stress experiments on Chlamydomonas reinhardtii} may relate more closely to a specific study description rather than the title or abstract.

After the fine-tuning phase, each ARC document from the dataset is divided into chunks and embedded separately using the fine-tuned ArcBERT model. These embeddings are then stored in an FAISS index, enabling retrieval at the whole document and individual chunk levels. This dual indexing strategy enhances precision and interpretability, allowing users to see which layers of ARC metadata are most relevant to their query. Previous semantic search systems have employed a similar approach to improve retrieval granularity and enhance user trust in model outputs~\cite{DBLP:conf/emnlp/KarpukhinOMLWEC20}. To support this method, each embedding is stored along with the information presented in Table~\ref{tab:chunk}, which is utilized to provide context-aware highlights in the retrieval interface and to aid in filtering and ranking based on the type of matched content. Indexing the documents and structures can be extended to include any new ARC metadata generated.
\begin{table}
\caption{Chunk data structure format for semantic indexing.}
\label{tab:chunk}
\begin{tabular}{|lll|}
\hline
\textbf{Field}        & \textbf{Type}    & \textbf{Description}                                                                                                                               \\ \hline
chunk\_text  & String  & \begin{tabular}[c]{@{}l@{}}The extracted chunk of data from an \\ ARC document (e.g., a study description \\ or assay name).\end{tabular} \\ \hline
field\_type  & String  & \begin{tabular}[c]{@{}l@{}}Field name in the ARC structure.\\  (e.g., study, assay, person, publication)\end{tabular}                     \\ \hline
arc\_id      & String  & \begin{tabular}[c]{@{}l@{}}A unique identifier that links the chunk \\ to its parent ARC document.\end{tabular}                           \\ \hline
chunk\_index & Integer & \begin{tabular}[c]{@{}l@{}}Position of the chunk in the data field \\ (used for ordering in case of multiple entries).\end{tabular}       \\ \hline
embedding    & Vector  & Dense representation computed by ArcBERT.                                                                                                 \\ \hline
\end{tabular}
\end{table}
\subsubsection{Hybrid Scoring}
Matching keywords from user queries can be highly valuable in scientific data exploration, especially with structured or formal metadata. This is particularly important for input queries that are short, partially structured, or composed solely of technical terms rather than complete natural language sentences. To address this, we leverage a hybrid scoring model that balances semantic and lexical similarity~\cite{Scoring}. The final relevance score is derived by combining normalized cosine similarity based on FAISS with BM25 scores. Both scores are scaled between 0 and 1 before being merged using a weighting strategy. This scoring model is utilized for both complete document retrieval and partial chunk retrievals as demonstrated in equation \ref{eq:hscore}.
\begin{equation}
    S_{\text{final}} = \alpha \cdot S_{\text{D}} + (1-\alpha) \cdot S_{\text{cmax}}
    \label{eq:hscore}
\end{equation}
where $S_{\text{D}}$ represents the weighted similarity score between the query vector and the ArcBERT document embeddings, while $S_{\text{cmax}}$ denotes the maximum weighted score among all chunks. The weight parameter, $\alpha$, can be adjusted to prioritize either the document or chunk scores. Additionally, the final score $S_{\text{final}}$ can be boosted when the user specifies certain metadata structure filters, such as investigation titles, assays, or studies.
\subsubsection{Post-Retrieval Summary}
The system offers a flexible interface, allowing users to define natural language queries interactively. The query processing layer generates query embeddings, performs semantic matching, and ranks the top-5 results using the hybrid scoring model. The retrieved documents are summarized in natural language, helping users understand the content with a coherent overview, which enhances usability. In the final stage of the retrieval process, we integrate Google's Gemini API to create concise, structured summaries for each top-ranked result (top-k). These summaries are designed solely to aid in interpreting the results and are not intended to influence their ranking.
\subsubsection{Metadata Updates}
As we refine the ArcBERT model using the latest snapshot of the ARC Metadata corpora from the Registry application, newly created ARC documents can be processed directly into document and chunk embeddings for indexing. The system will continue to incorporate the new ARCs for semantic matching. However, the update process must be streamlined to ensure the model is fine-tuned at regular intervals with the most up-to-date metadata corpora.
\section{Experiments}\label{sec:experiments}
We assess the effectiveness of ArcBERT for metadata exploration compared to the traditional text search engine, ElasticSearch. Our main objective is to evaluate the relevance of the top-5 retrieved results and their ranking. 
\subsection{Setup}
All experiments were conducted on a dedicated machine running Ubuntu 22.04.5 LTS. The server is equipped with an AMD EPYC 7662 64-Core processor and 1TB of RAM. Model training tasks were accelerated using an NVIDIA GPU, specifically A100 PCIe with 80 GB of memory. The development environment is based on Python 3.10, utilizing PyTorch 2.0 as the primary deep learning framework.

For sentence-level semantic retrieval, we implemented the sentence-transformers library (v2.2.2), which is built on Hugging Face Transformers. We used FAISS (v1.7.3) for approximate nearest neighbor search in high-dimensional space, while ElasticSearch (v8.9.1) served as the baseline keyword-based retrieval engine.

ArcBERT, based on the all-mpnet-base-v2 architecture, is trained on structured ARC metadata. It was optimized using the Multiple Negatives Ranking Loss (MNRL) function, which promotes semantic coherence in the embedding space. Additionally, the Gemini summarization module was accessed via the official Google Generative AI client, utilizing the Gemini-1.5-flash model.
\subsection{Benchmark Query-set}
We have curated a set of 122 test queries to assess the effectiveness of ArcBERT compared to ElasticSearch. These queries are logically organized into nine distinct categories, each representing different user query requirements. The queries range from simple keyword searches (abbreviated as KW) to more complex inquiries that incorporate hierarchical layers of metadata, such as parameters (KWPAR), studies (KWSUD), assays (KWA), or simply queries in natural language (SWK and SEM). This query set is designed to simulate real-world scenarios where users might approach data exploration from varied perspectives—whether by seeking specific experimental techniques or referencing known contributors or publications. An example query includes
\begin{itemize}
    \item Semantic: \emph{Which ARC investigates how disease resistance genes are constantly active in crops like tomatoes and potatoes?}
    \item With keywords: \emph{Which ARC study about \textbf{global expression patterns} of \textbf{R-genes} in tomato and potato?}
\end{itemize}
\subsection{Query Execution}
Each query is executed separately on both retrieval systems, with the top five results selected based on matching scores and averaged for each query category. Notably, including structured fields such as people, parameters, assays, and studies enables hierarchical filtering, allowing users to refine their search results further. These filters mimic how users may narrow their queries when exploring complex metadata. 

We compare each result set averaged for respective query categories with multiple retrieval metrics, including mean rank, mean reciprocal rank (MRR), and similarity scores between the query and the results. These metrics assess the precision of ranking and the semantic alignment of the results. MRR highlights the importance of retrieving relevant documents early, while mean rank provides a broader view of how results are distributed among the top ranks.

A consistent normalization strategy for the scoring values is essential to ensure meaningful comparisons between two retrieval systems. ElasticSearch utilizes the BM25 ranking algorithm, which produces unbounded output scores that can vary significantly based on the frequency and distribution of query terms within the indexed corpus. Without normalization, these variations could lead to misleading quantitative assessments, particularly when comparing performance across shared metrics like Mean Reciprocal Rank (MRR) or evaluating the quality of ranked outputs. This approach is consistent with evaluation methods used in large-scale information retrieval (IR) tasks such as TREC and CLEF, where different systems are benchmarked using shared ranking-based metrics instead of relying on their native scoring outputs. To achieve this, we apply a global min-max normalization strategy for the ElasticSearch scores, as described below:
\begin{equation}
    \text{Normalized Score}_{\text{ES}} = \frac{S - S_{\text{min}}}{S_{\text{max}} - S_{\text{min}}}
\end{equation}
\begin{figure}
    \centering
    \includegraphics[width=\columnwidth]{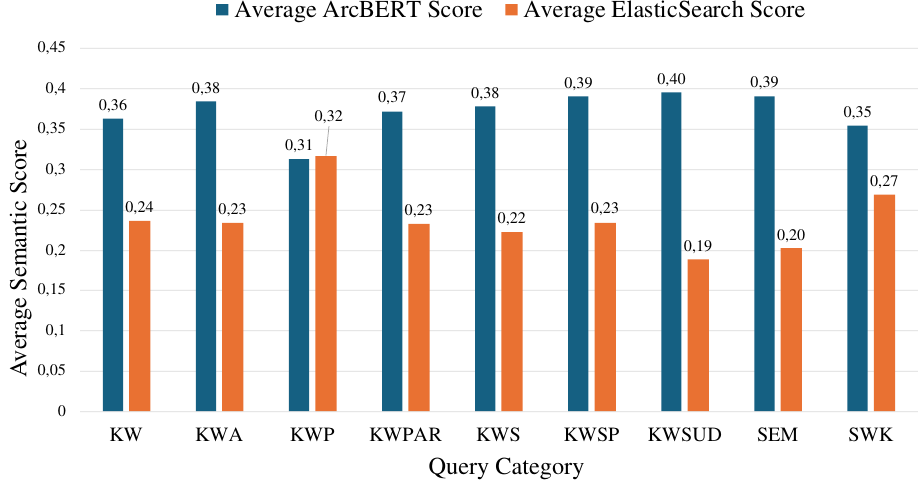}
    \caption{Average top-5 similarity scores across query categories for ArcBERT and ElasticSearch.}
    \label{fig:scores}
\end{figure}
\subsection{Results}
\subsubsection{Average Retrieval scores}
Figure.~\ref{fig:scores} demonstrates that ArcBERT consistently achieves higher average similarity scores for the top-5 results across a wide range of query categories. These results indicate that language models like ArcBERT offer significant advantages in understanding semantically rich and natural language queries, allowing them to deliver result sets more closely aligned with user intentions. While ElasticSearch efficiently retrieves the best matching top result for queries containing specific keywords, it becomes less effective when considering the average scores for the top-5 results. This finding is expected, as token-based retrieval systems typically perform better when user queries are closely aligned with indexed metadata fields. Interestingly, ArcBERT's hybrid approach—which utilizes cosine similarity and lexical BM25 scoring—enhances consistency across various query categories for the top-5 results.
\subsubsection{Mean top rank scores}
\begin{figure}
    \centering
    \includegraphics[width=\columnwidth]{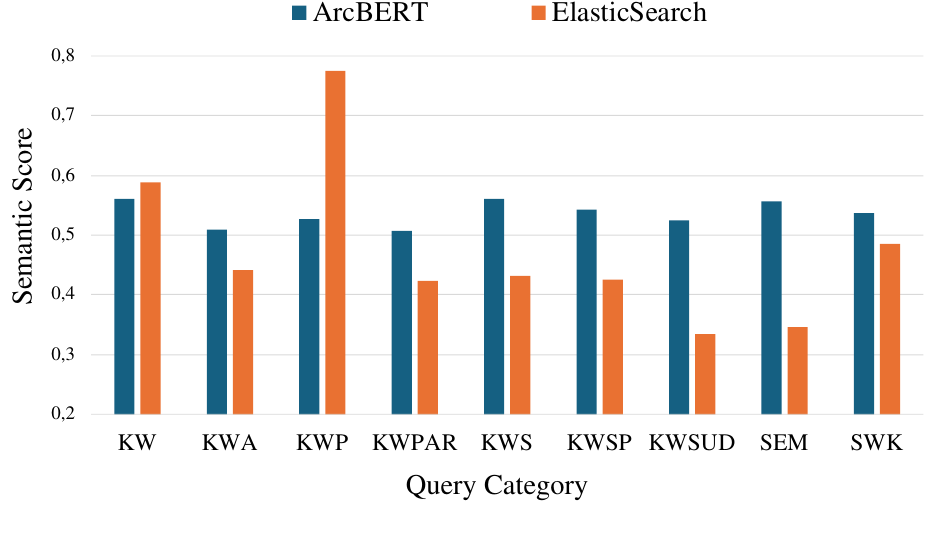}
    \caption{Mean similarity score of the top-1 result per query category.}
    \label{fig:meantopscore}
\end{figure}
Figure.~\ref{fig:meantopscore} illustrates the average score of the top-ranked retrieved results for ArcBERT and ElasticSearch across different query categories. ArcBERT consistently achieves significantly higher first-rank scores in categories where the queries do not show substantial lexical overlap with ARC metadata, thanks to its semantic encoding capabilities. In contrast, ElasticSearch struggles in these categories with first-rank scores. However, ElasticSearch performs comparably or even better in keyword-heavy categories such as KW, KWP, and KWA, where it benefits from exact token matches.
\subsubsection{Mean reciprocal ranks}
\begin{figure}
    \centering
    \includegraphics[width=\columnwidth]{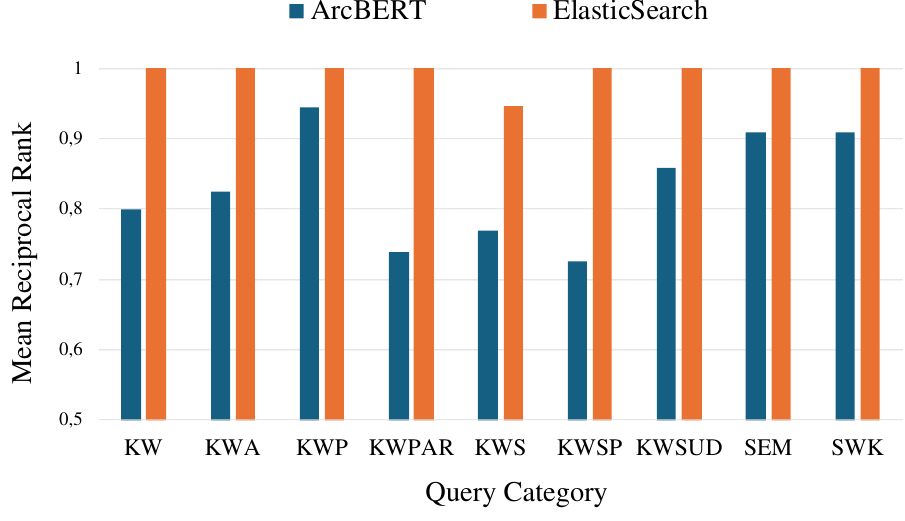}
    \caption{MRR comparison across the query categories.}
    \label{fig:mrr}
\end{figure}
The Mean Reciprocal Rank (MRR) is a metric that measures how well a system ranks the highest expected result within a list of retrieved documents. Figure.~\ref{fig:mrr} compares the MRR across different query categories. Unsurprisingly, ElasticSearch outperforms ArcBERT in all categories, demonstrating its effectiveness in ranking the best-matching documents at the top. In contrast, the ArcBERT scoring model focuses more on consistent semantic matching.
\section{Conclusion \& Future Work}
In this paper, we introduced ArcBERT, a semantic search engine based on Sentence-BERT, designed to help users explore omics metadata through natural language queries. ArcBERT is pre-trained and fine-tuned to comprehend plant science terminology and the various layers of integrated omics metadata. We evaluated the effectiveness of ArcBERT by comparing it to a traditional text search engine using a query dataset that simulates real-world user query patterns. As anticipated, the results of our experiments show that ArcBERT can understand natural language-style queries and the structure of the metadata, outperforming text search engines in retrieving consistent results that are semantically closer to user queries. However, we also observed that text search engines are still effective at retrieving the best matching results through keyword matching.

ArcBERT offers a significant advantage in usability, as it can comprehend natural language queries. The option for users to write queries in plain English is invaluable in the bioscience domain, where keyword-centric data retrieval plays an important role. As a direction for future work, we plan to fine-tune its hybrid scoring function to better balance semantic and keyword query matching. Additionally, we aim to streamline the process of retraining the model with new metadata before integrating ArcBERT into the existing ARC Metadata Registry application.
\section*{Acknowledgment}
We acknowledge the support of DataPLANT (NFDI 7/1 – 42077441) as part of the German National Research Data Infrastructure funded by the Deutsche Forschungsgemeinschaft (DFG – German Research Foundation).

\bibliographystyle{IEEEtran}
\bibliography{paper}
\end{document}